\documentclass{article}

\usepackage{ieee_tns}
\usepackage{epsf}   
\usepackage{multicol}
\usepackage{graphics}
\usepackage{graphicx}

\begin{document}

\begin{center}
\title{Design and Operation of Front-End Electronics for the HERA-B Muon
Detector}

\author{M.~Buchler, W.~Funk, A.~Gutierrez, R.F.~Harr, P.E.~Karchin, \\
P.~Liu, S.~Nam, J.G.~Shiu, S.F.~Takach \\ Department of Physics, Wayne State
University, \\ Detroit, MI 48201 \\ \vspace*{0.5cm}
M.~Atiya, D.~Padrazo \\ Dept. of Physics, Brookhaven National Laboratory, \\
Upton, NY 11973 \\ \vspace*{0.5cm}
A.~Arefiev, S.~Barsuk, F.~Khasanov, L.~Laptin, V.~Tchoudakov, I.~Tichimirov,
M.~Titov, Y.~Zaitsev \\  
Institute of Theoretical and Experimental Physics (ITEP), \\ 117259 Moscow,
Russia}  
\end{center}
\maketitle

%---   cr.notice.tex
%
%---   Stephen F. Takach --- 23 November 1998.
%      This is a \input file for paper.tex to generate a copyright notice for
%      publishing the ieee paper on a preprint server or on a webpage.. This
%      material came from the IEEE website. 

\twocolumn[
\vspace*{10.0cm}
\begin{center}
  \large
  
  \copyright 1998 IEEE. Personal use of this material is permitted.  However,
  permission to reprint/republish this material for advertising or promotional
  purposes or for creating new collective works for resale or redistribution to
  servers or lists, or to reuse any copyrighted component of this work in other
  works must be obtained from the IEEE.
  \vspace*{1.0cm}

  This material is presented to ensure timely dissemination of scholarly and
  technical work. Copyright and all rights therein are retained by authors or
  by other copyright holders. All persons copying this information are expected
  to adhere to the terms and constraints invoked by each author's copyright. In
  most cases, these works may not be reposted without the explicit permission
  of the copyright holder.
\end{center}
]

\newpage
\maketitle

\begin{abstract}
  
We have implemented a cost-effective design for the readout electronics of
both the anode wires and the cathode pads of large area proportional wire
chambers for the HERA-B muon system based on the ASD-08 integrated circuit.
To control and monitor the large number of readout channels, we have built a
distributed control system based on Philips Semiconductors' I$^{2}$C bus and
microcontrollers. To date we have installed about 10800 channels of muon
chambers and electronics. The average single channel noise occupancy is less
than $10^{-5}$, and the detectors have been operated with target
interaction rates as high as 70~MHz.

%Initial indications are that the noise performance
%is good and that the system meets the high rate needs of HERA-B.

\end{abstract}

\section{Introduction}
   
The HERA-B experiment at DESY is designed to observe and measure CP violation
in B decays. Precise CP violation measurements will either provide an accurate
determination of the values of the CKM matrix or will point to new physics
beyond the Standard Model. In the HERA-B experiment, HERA's 820 GeV circulating
proton beam interacts with a wire target placed in the beam's halo. The
proton-nucleus interactions copiously produce B events which must be culled
from the large non-B background. This places great importance on triggering and
event selection. For B decays, this means identification of non-prompt J/$\psi$
decays and leptons with high transverse momentum.  Hence, the muon detector
plays a primary role in particle identification and triggering in the
experiment.

The muon system is segmented into 4 large superlayers, which measure roughly
6~m~$\times$~8~m.  The 4 superlayers are interleaved with iron-loaded concrete
to screen out hadrons, and they are located in the region from about 16~m to
20~m downstream of the target. The system covers angles from about 10~mrad near
the beam pipe up to maximum angles of 160~mrad vertically (in y) and of
220~mrad horizontally (in the bend plane or x-direction). The entire system
consists of about 32000 channels of readout. Figure \ref{fig:MuonSystem} shows
a view of the entire muon system.
\begin{figure}[hbtp]
\begin{center}
%  \resizebox{2.5in}{!}{\includegraphics[angle=-90,bb=98 356 669 788]{iso_view.ps}}
%\includegraphics[angle=-90,width=1.5in,keepaspectratio=true,bb=62 219 633
%651]{iso_view.ps}   
%\centerline{\epsfxsize 3.5in \epsffile{iso_view.ps}}
  \vspace*{2.5in}
  {\caption{\label{fig:MuonSystem} \protect \small The HERA-B Muon System. This 
      view shows the superlayers interleaved with the absorber.}} 
\end{center}
\end{figure}

Three kinds of detectors make up the muon system: tube chambers, pad chambers,
and pixel chambers. The tube and pad chambers are closed-cell proportional wire
chambers. The cell dimensions are $14\times 12$~mm$^{2}$. About 15000 anode
wires and 8000 cathode pads in these chambers cover angles from about 50~mrad
up to the maximum angles in x and y. The first two superlayers consist of tube
chambers arranged into 3 stereo views. The wire angles are $\pm 20^{o}$ and
$0^{o}$ with respect to the vertical. Superlayers 3 and 4 consist of pad
chambers arranged only into the $0^{o}$ view. Each superlayer also has a pixel
system that makes up about 9000 channels in the muon system. The pixel chambers
cover the small angles from 10-50~mrad near the beam line.

The tube and pad chambers each have two layers of 16 anode wires of pitch
16~mm; one layer is offset by 1/2 cell from the other layer. The chamber width
is 262~mm, and the lengths vary; a typical length is approximately 3~m. The
anode wires are made of gold-plated tungsten of diameter 45~$\mu$m. Inside each
pad chamber are two copper clad phenolic boards. Sixty cathode pads (2 columns
of 30) of approximate size $130\times 100$ mm$^{2}$ are cut into the copper
cladding of each phenolic board. The boards are mounted inside the pad chamber
so that the pads in the two layers are aligned. Figure \ref{fig:Mu1.Vertical}
shows superlayer 1, the first completed superlayer, before insertion into the
muon absorber. For more information on the construction and geometry of the
muon chambers, see references \cite{herab:proposal} and \cite{herab:tdr}.
\begin{figure}[hbtp]
\begin{center}
%  \resizebox{\textwidth}{!}{\includegraphics {Mu1.Vertical.eps}}
%\centerline{\epsfxsize 3.5in \epsffile{Mu1.Vertical.eps}}
\vspace*{3.5in}
  {\caption{\label{fig:Mu1.Vertical} \protect \small Muon Superlayer 1. This is
      a view of tube chambers mounted on the support frame before insertion
      into the muon absorber. The two halves of the frame are slightly
      separated. the $\pm 20^{o}$ stereo view chambers are visible. In the
      upper right corner, the control cards and low voltage bus bars are
visible.}}
\end{center}
\end{figure}

%---   paper.karchin.tex
%
%---   Paul E. Karchin --- 12 August 1998.
%
%---   Stephen F. Takach --- 14 August 1998.
%      Put into latex format to act as \input for paper.tex.

\section{Front End Electronics}

%Because of the large physical extent of the tube and pad chambers, the
%electronics is necessarily distributed over a large area.

%There are approximately 15000 tube channels and 8000 pad channels.

%pixel chambers - 9000 channels 

\subsection{Choice of Front End Device}

The HERA-B spectrometer contains several large systems requiring high gain,
front-end amplifiers with discriminated (yes/no) readout. These systems are the
large angle drift chamber tracking system (96000 channels), the muon detector
(32000 channels), the detectors for the high transverse momentum trigger
(26000 channels), and the ring imaging Cherenkov (RICH) detector (28000
channels).  The large number of channels dictated the need for a cost effective
solution that would be usable in all systems. All the detectors mentioned
above, except for the RICH phototubes, employ gas avalanche techniques. The
high bunch crossing rate at HERA-B (10 MHz) makes desirable the highest
possible electronic gain in the front end amplifiers. This allows these
detectors to operate at the lowest possible avalanche gain.

An attractive solution is the ASD-08 integrated circuit \cite{Newcomer:asd08}
developed for the straw tube drift chamber tracker for the Solenoid Detector
Collaboration at the Superconducting Super Collider. In the muon tube chambers,
for example, at nominal high voltage of 2.3~kV, a minimum ionizing particle
produces an anode signal with a mean value of about 30~fC within the 7~ns
shaping time of the ASD-08. Using a digital storage oscilloscope, this was
determined by direct measurement of anode pulses from cosmic rays. This matches
perfectly with the fC level thresholds which the ASD-08 allows. Packaged chips
were produced in large quantities for HERA-B in 1997 for \$2.34 per channel,
not including testing.  The wafers were fabricated by MAXIM Microelectronics
(Beaverton, Oregon) and packaged by Hana Technologies (Hong Kong).

\subsection{Anode Wire Front End Electronics}
\label{subsec:anodewirefee}

The front end electronics for reading out the anode wires are implemented on
32-channel, printed circuit boards (``tube cards'', for short). The size is
roughly $200 \times 80$ mm$^{2}$. They are mounted in close proximity to the
anode wire feedthroughs of the tube and pad chambers. Figure
\ref{fig:tubecardmounted} shows one such card mounted on a tube chamber. Each
anode wire is connected to the card via a short (approx. 6 cm), insulated wire.
The wire is rated for 18 kV (from Reynolds Industries, Los Angeles,
California).  It is soldered to the feedthrough at one end and has a single pin
connector at the card end.  The wire is kept as short as possible to minimize
electromagnetic pickup to the amplifier input.
\begin{figure}[hbtp]
\begin{center}
% \resizebox{3.0in}{!}{\includegraphics 
%   {tubecardmounted.eps}} 
%\centerline{\epsfxsize 3.5in \epsffile{tubecardmounted.eps}}
\vspace*{3.5in}
{\caption{\label{fig:tubecardmounted} \protect
      \small The Tube Card Mounted on a Chamber. The flexible wires from the
      anodes to the ASD card input are visible. The signal, power, and high
      voltage cabling to the tube card and the walls of the rf shield box are
      also visible. The copper foil and the top cover to the rf shield are
      not installed.}}
\end{center}
\end{figure}

%HV on same board as signal electronics to save manual labor

% A photograph of the component side of a tube card is shown in Figure yy.

\begin{figure}[hbtp]
\begin{center}
%  \resizebox{\textwidth}{!}{\includegraphics 
%    {tuberead.eps}} 
\centerline{\epsfxsize 3.5in \epsffile{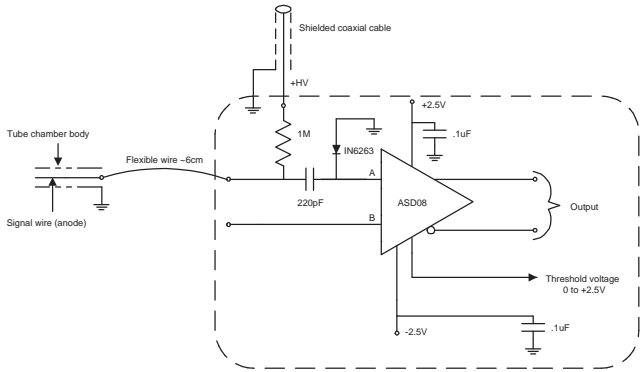}}
%\vspace*{3.5in}
{\caption{\label{fig:tubecardschem} \protect
      \small A Schematic of the Readout from Anode Wire to ASD Input. The
schematic shows 1 of 32 channels on the tube card. The dashed line surrounding
the ASD represents the boundary of the tube card.}}
\end{center}
\end{figure}

A schematic diagram of the input stage from anode wire to ASD-08 appears in
Figure \ref{fig:tubecardschem}.  The high voltage distribution resistors
(1~M$\Omega$) are located on the same PCB as the ASD-08 package.  The high
voltage is provided via an HTC-50-1-1, miniature coaxial cable.  This flexible
cable has an outer diameter of 3.3 mm.  An internal discharge shield allows
operation up to 5 kV.  The high voltage is blocked from the ASD-08 by a 220~pF
ceramic, wire-lead capacitor rated at 6 kV (XICON Passive Components,
Arlington, Texas). All other components are of the surface mount type. To
protect against air discharges, the board is spray coated with a clear
insulating varnish, Krylon 7001.  A Schottky-barrier diode, type 1N6263,
protects the ASD-08 if a discharge occurs in the chamber between the anode wire
and ground.  In this case, the capacitor discharge current will flow through
the diode.

The design of the tube card includes a number of features to reduce
electromagnetic pickup of noise and to increase the impedance of feedback
paths. Input and output traces of the ASD-08 are sandwiched between ground and
power planes of the circuit board.  Overall, the printed circuit board has 5
layers. The ASD-08 has differential inputs; one is connected to an anode wire,
and the other is connected to a parallel trace equal in length. This allows for
differential cancellation of noise pickup along the input trace path.
Decoupling capacitors (0.1 $\mu$F) are located near the ASD-08 package at the
bias pins.  Also, 10 $\mu$F capacitors filter the bias voltages (-2.5 V and
+2.5 V) near the power input connector.

The ASD output drives long (up to 20~m), twisted-pair signal cables. The signal
cables plug into a module termed the Front End Driver (FED) by HERA-B. It is a
custom module built commercially for HERA-B by MSC (Stutensee, Germany). The
FED pulls up the differential output lines of the ASD by terminating them
through 62~$\Omega$ into +1.6~V. The FED latches the ASD data and holds it in a
128 event deep buffer while the pretrigger and the first level trigger decide
whether the data is worthy of sending to the second level trigger.

\subsection{Cathode Pad Front End Electronics}

The large length of the muon pad chambers ($\sim 3$~m) and need to transport
the cathode signals over large distances requires that the front end
electronics be separated into two stages. The first stage is a preamplifier. It
is mounted near the cathode pads, inside the pad chambers, and on the side of
the phenolic board opposite to the anode wires. Each preamplifier board is
roughly $30~\times~25$~mm$^{2}$ and provides preamplification for 3 cathode
pads. Two bus lines cut into the copper cladding of the phenolic boards supply
power (-3.5~V) and ground to the preamplifiers.

\begin{figure}[hbtp] 
\begin{center}
%  \resizebox{\textwidth}{!}{\includegraphics
%    {preamp.eps}} 
\centerline{\epsfxsize 3.8in \epsffile{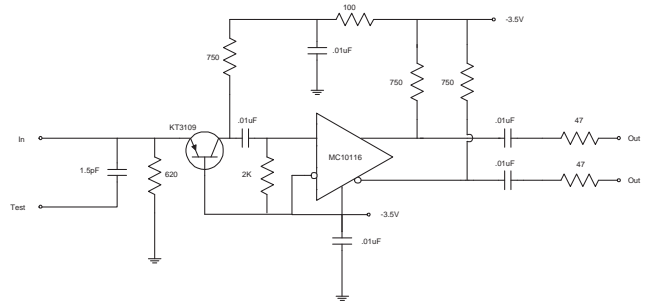}}
%\vspace*{3.8in}
{\caption{\label{fig:preamp} \protect \small Schematic
      Diagram of One Channel of the Cathode Pad Preamplifier. }}
\end{center}
\end{figure}

\begin{figure}[hbtp] 
\begin{center}
%  \resizebox{\textwidth}{!}{\includegraphics 
%    {pad.eps}} 
\centerline{\epsfxsize 3.7in \epsffile{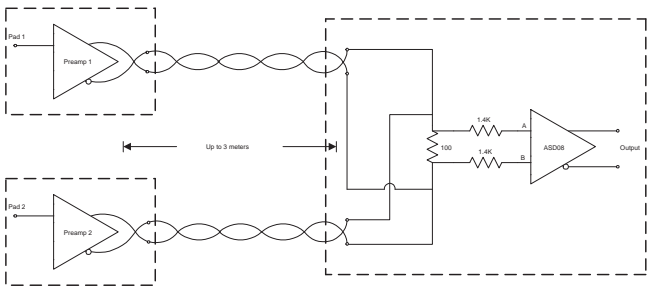}}
%\vspace*{3.7in}
{\caption{\label{fig:padcard} \protect \small The Readout of the
Cathode Pads from the Preamplifiers to the Input of the ASD. The dashed line
surrounding each preamplifier is the boundary of the cathode pad. The dashed
line around the ASD circuitry is the boundary of the pad card. The outputs of
two preamps are OR'd together at the input of the pad card.}}
\end{center}
\end{figure}

The schematic for one such preamp channel appears in Figure
\ref{fig:preamp}. The input stage of the preamplifier circuit consists of a
KT3109 transistor in common base configuration. This allows a small input
impedance to the preamplifier circuit. The output stage consists of an ECL Line
driver (MC10116 by Motorola). It drives long ($\sim 3$~m) twist-and-flat cables
from near the cathode pad to the input of the readout electronics located at
the end of the chamber. The overall gain of the preamplifier is approximately
1.2~mV/$\mu$A. To reduce the number of readout channels, a pad from one
phenolic board is OR'd with a pad from the other board. Figure
\ref{fig:padcard} shows the OR-ing of the two pads.

The cathode pad front end card (``pad card'') is implemented as a 5-layer
printed circuit board of dimensions $238 \times 80$~mm$^{2}$. Figure
\ref{fig:pad_card_photo} shows the top and bottom sides of the pad card. The
input portion of the pad card contains a termination and voltage divider
network. The network terminates the cables and attenuates the large signal from
the preamplifier into the range expected by the ASD. This allows us to use the
ASD integrated circuit in this application. After this network, the inverting
and noninverting outputs from the driving stage of the preamplifier feed into
the A and B inputs, respectively, of the ASD-08.

\begin{figure}[hbtp]
   \begin{center}
%   \centerline{\epsfxsize 3.5in \epsffile{pad_card_photo.ps}}
   \vspace*{3.5in}
   \caption{\label{fig:pad_card_photo} \protect \small The Cathode Pad Readout
     Card. The upper portion of the photo shows the top side of the card with
     the ASD chips mounted center left. The input connectors to the ASDs are on
     the upper left. The output connectors from the ASDs are on the lower left.
     The power and control connector is on the lower middle, and the connectors
     for the preamp power are on the upper and lower right. The lower portion
     of the photo shows the back side of the card on which are mounted the DAC
     chips.}
   \end{center}
\end{figure}

\subsection{Test and Calibration Procedures}
\label{subsec:testandcalib}

We receive untested ASD-08 chips from the manufacturer. Our first task is to
test them before mounting them on boards. For this test, channel thresholds are
set to 360~mV, and through a 1 pF capacitor, each channel receives an input
pulse of approximately 8 fC of charge. The output from the ASD goes to a custom
receiver module via twisted pair cables. The receiver module provides the
pullup voltage to the ASD output lines, and it converts the ASD signals to NIM
levels for observation on an oscilloscope. Under these conditions, a channel is
``good'' if it gives a pulse of 10-20~ns in width with few noise pulses. About
75\% of the chips pass this initial test.

\begin{figure}[h]
   \begin{center}
%      \resizebox{\textwidth}{!}{\includegraphics
%    {AsdCardTestStand.eps}} 
\centerline{\epsfxsize 3.5in \epsffile{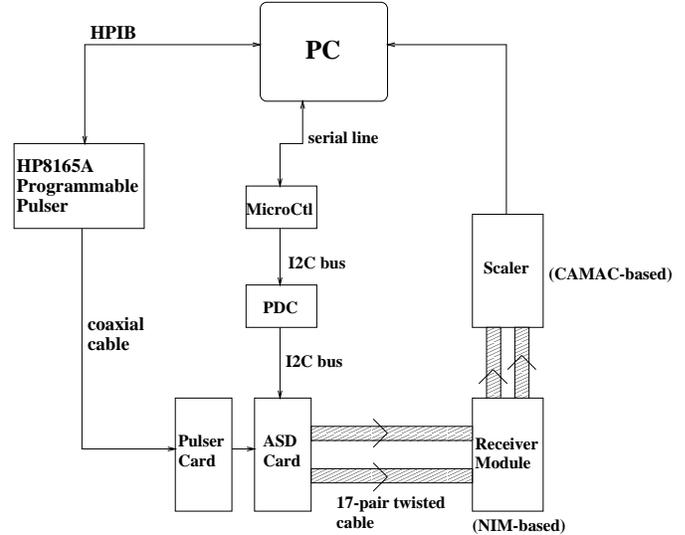}}
%\vspace*{3.5in}
{\caption{\label{fig:AsdCardTestStand} \protect
\small Block Diagram of the Test Stand for the ASD Cards. The microcontroller
(MicroCtl), the power distribution card (PDC), and the I$^{2}$C bus are
discussed in Section \ref{sec:ctlsys}.}}
   \end{center}
\end{figure}

After assembly, each board undergoes a test and calibration of all 32 channels.
We constructed a test stand to automate the process. The test stand consists of
the apparatus shown in Figure \ref{fig:AsdCardTestStand}. The basic idea of the
test is to measure the efficiency of a channel as a function of 2 parameters,
{\em injected charge} and {\em threshold voltage}. A test program on the PC
controls a pulser (HP8165A) through an HPIB interface. The program varies the
height of a rectangular pulse that the pulser injects into a passive, RC
differentiator network; this varies the spike of charge injected into the ASD.
Our custom receiver module converts the ASD output to ECL levels which go to a
CAMAC-based scaler for counting.

\begin{figure}[h]
   \begin{center} 
%   \resizebox{\textwidth}{!}{\includegraphics {calib-pad.eps}}
\centerline{\epsfxsize 3.5in \epsffile{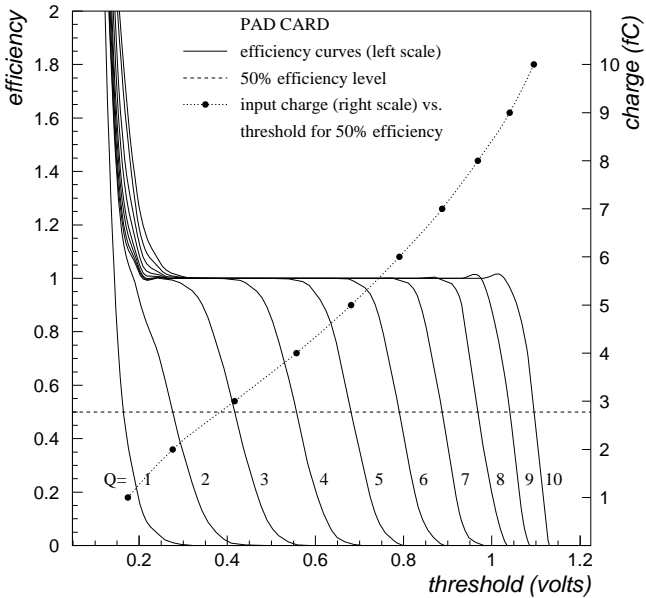}}
%\vspace*{3.5in}
   {\caption{\label{fig:calib-pad} \protect \small Efficiency Curves (solid
   lines) and the Calibration Curve (dashed line) for One Channel of a Pad
   Card. The text (Section \ref{subsec:testandcalib}) contains an explanation
   of the calibration curve.}}   
   \end{center}
\end{figure}

Testing is done in two parts to avoid cross-talk between neighboring channels.
For a given pulse height and threshold voltage, the pulser sends 10$^{5}$
pulses to the 16 odd channels on the ASD card. The efficiency is the number of
pulses counted by the scaler divided by the 10$^{5}$ injected pulses. The test
program scans the efficiency at 10 different pulse heights before changing the
threshold. After the odd channels are tested, the operator manually switches
the input pulse to the even channels so that they can be tested in the same
way. Figure \ref{fig:calib-pad} shows the efficiency curves for a good channel
of a pad ASD card.

The last step in the testing procedure is to extract the calibration curve from
the efficiency data. The filled circles in Figure \ref{fig:calib-pad} are the
data derived from the efficiency curves. They are the values of the threshold
voltage for which the channel is 50\% efficient at each value of the injected
charge. The test program fits this data to the form: $V_{th} = a + b\tanh (c
Q)$.  $Q$ is the injected charge, and $V_{th}$ is the threshold voltage at
which the channel is 50\% efficient; the dotted line shows the calibration
curve. As a final step for each channel, the program saves the fit parameters
$a, b, c$ in an EEPROM located on the ASD card. This allows for setting the
threshold in terms of a charge threshold rather than a voltage threshold.

The final tests of the ASD cards are done after mounting them onto the
chambers. The first test is simply to reduce the threshold enough so that the
channel is on 100\% of the time. The second test is to turn off the channel by
increasing the threshold. This test relies on the highly nonlinear behavior of
the threshold above threshold voltages of about 1.1-1.2~V. The charge input
required for 50\% efficiency increases dramatically for voltages above these
values. The channel can easily be switched off by setting the threshold voltage
high enough. Lastly, a working channel must have a noise level less than a
few~$\times$~10$^{-4}$ in the neighborhood of a threshold of 0.8~V (about 4 fC
injected charge).

%---   paper.takach.tex
%
%---   Stephen F. Takach --- 14 August 1998.
%      For \input into paper.tex.
%

\section{Power Distribution and Control System}
\label{sec:ctlsys}

\subsection{General Design Overview}

With so many front end channels in the muon system, monitoring and controlling
them is an important consideration in properly running the apparatus. Providing
one threshold voltage per card (32 channels) would be easier than segmenting
the threshold voltages further.  However, the optimum threshold voltage varies
with the ASD channel. The way to maintain the most uniform hit detection
efficiency is to provide one threshold voltage per ASD channel.  Furthermore,
control of individual thresholds allows for setting patterns of on or off
channels for trigger testing. Therefore, each front end electronics card
contains one 8-channel digital-to-analog converter (DAC) per ASD-08 chip, for a
total of 4 DACs per electronics card. To store default threshold values and the
threshold calibration per channel, each front end electronics card also
contains an EEPROM.

\begin{figure}[hbtp] 
\begin{center}
%  \resizebox{\textwidth}{!}{\includegraphics 
%    {SlowCtlHierarchy.eps}} 
\centerline{\epsfxsize 3.5in \epsffile{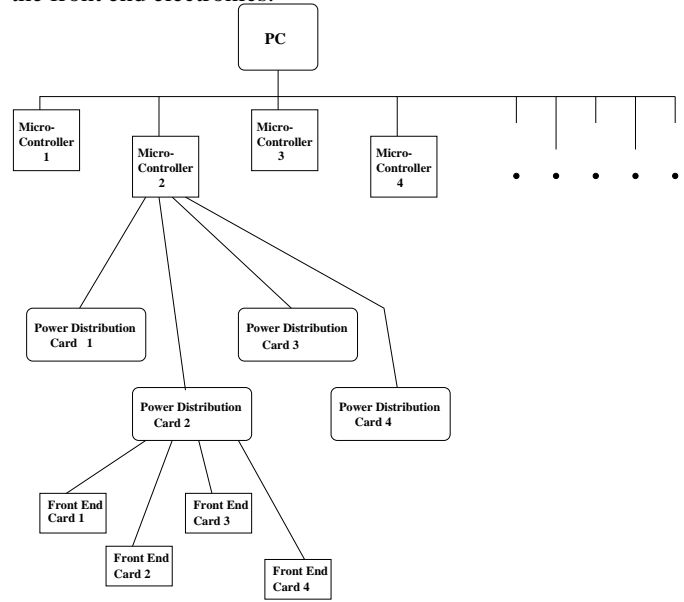}}
%\vspace*{3.5in}
{\caption{\label{fig:SlowCtlHierarchy} \protect
\small The Hierarchy of the Control System.}}
\end{center}
\end{figure}

The front end cards (ASD cards) and all their control electronics form a
hierarchy from a computer at the highest level to the front end cards at the
lowest level, as evident from Figure \ref{fig:SlowCtlHierarchy}. On the test
bench, at the highest level is a PC running under the Windows OS. In the
experiment, the highest level is a VME-based Cetia CPU Module, running under
the Lynx OS. The computer communicates with a given network of {\em
microcontroller cards} through a serial port connected to roughly 50~m of
cable. Attached to each microcontroller card by cables of approximately 1-3
meters in length are up to 4 {\em power distribution cards}. Each power
distribution card fans out low voltage power to a maximum of 4 front end cards
and acts as a control gateway to them. Therefore, a single microcontroller card
controls up to 16 front end cards. Because the muon system is so large, the
microcontroller and power distribution cards are distributed along the support
frame, near the front end electronics.

\subsection{The Microcontroller Card}

The intelligence in the control system is located in 2 places: near the
front end cards and near the operator. The microcontroller card, Figure
\ref{fig:uCtlCard}, is mounted on the frames near the front end electronics and
acts as the local intelligence in the system. The kernel of the microcontroller
card is the CMOS 80C652 8-bit microcontroller chip from Philips Semiconductors
\cite{philips:uCtlr}.
\begin{figure}[hbtp] 
\begin{center}
%  \resizebox{\textwidth}{!}{\includegraphics 
%    {uCtlCard.eps}} 
%\centerline{\epsfxsize 3.5in \epsffile{uCtlCard.eps}}
\vspace*{3.5in}
{\caption{\label{fig:uCtlCard} \protect
\small The Microcontroller Card. The dimensions of the card are approximately
$140 \times 90$ mm$^{2}$. The serial line connector is at the top of the
figure, and the 4 connectors to the power distribution cards are at the
bottom.}} 
\end{center}
\end{figure}

The microcontrollers can use either internal or external memory. We have
configured them to run using both external program memory and external data
memory. The external program memory consists of a 32K~$\times$~8-bit CMOS EPROM
(CY27C256 by Cypress Semiconductor, San Jose, California). The external data
memory is a 32K~$\times$~8-bit CMOS RAM (CY62256 by Cypress). The program
memory is mounted in a socket on the printed circuit board to allow for removal
and reprogramming.

The microcontroller card has 2 communication interfaces with the world external
to it. To communicate with the PC or Cetia, it uses an RS422 (differential)
serial line.  The microcontroller has a single-ended, full-duplex UART built
into the chip. Off-chip, a 75ALS180 line driver interfaces the
microcontroller's serial port to the differential serial line.

The second communication interface to the outside world is the I$^{2}$C bus
\cite{philips:i2cbus}. The microcontroller chip has a built-in I$^{2}$C bus
interface. This allows it to communicate with the DAC and EEPROM integrated
circuits which also have built-in I$^{2}$C interfaces. The microcontroller is
the only I$^{2}$C master on the bus. To drive cables of several meters between
the microcontroller card and the power distribution cards, the microcontroller
chip is interfaced to the bus via the bipolar 82B715 I$^{2}$C bus driver
integrated circuit (by Philips). It buffers both the clock and the data lines
of the I$^{2}$C bus.

Another function of the microcontroller is to monitor the low voltage power
supplied to the front end cards.  Via a twist-and-flat cable plugged into the
microcontroller card, each PDC returns analog monitoring voltages for the
current and voltage levels on the $\pm 2.5$ V, +5 V, and -3.5 V power. The 8
monitoring voltages are multiplexed into a single 8-bit ADC. An
end-of-conversion (EOC) on the ADC is tied to the start signal, so that the ADC
continually samples its inputs. The EOC also interrupts the microcontroller so
that it will read the digitized ADC data into its memory.

The program running in the microcontroller handles the communication and
monitoring functions. During power-on or reset, the microcontroller initializes
the serial port and the ADC clock. It establishes the serial port interrupt
priority as the highest. Upon command from the PC or Cetia, the microcontroller
will initialize I$^{2}$C communications. Then it will enter a wait-loop. From
this state it will respond to interrupt requests from the serial port or from
the ADC end-of-conversion. Communications over the serial line are in the form
of character message strings which the microcontroller must first parse and
then act on. The PC or Cetia can send a message which asks the microcontroller
to write or read a piece of its data memory. What the microcontroller writes to
its memory may be data or a command. The memory is divided into data and
command registers, and if the microcontroller finds a command in the command
register, it will execute it.

\subsection{Power Distribution Card (PDC)}

In the control system hierarchy just between the microcontroller and the front
end cards are the power distribution cards, Figure \ref{fig:PDC}. They have
three main purposes. The first is simply the job for which it was named: to
distribute the low voltage power of $\pm 2.5$ V and +5V to the front end cards;
additionally, for the pad chambers, the power distribution card also
distributes -3.5 V to the pad preamps. The second purpose is to monitor those
low voltages. Each PDC uses op-amps to sense the currents and voltages.
\begin{figure}[hbtp] 
\begin{center}
%  \resizebox{\textwidth}{!}{\includegraphics 
%    {PDC.eps}} 
%\centerline{\epsfxsize 3.5in \epsffile{PDC.eps}}
\vspace*{3.5in}
{\caption{\label{fig:PDC} \protect
\small The Power Distribution Card. The card is mounted in an aluminum box to
shield it from rf pickup. The 4 connectors to the front end cards are at the
top of the figure. The connector to the microcontroller is in the lower right,
and the 4 connectors for the pad preamp power are on the left.}}
\end{center}
\end{figure}

The third and last task of the PDCs is to act as an I$^{2}$C gateway. The
microcontroller card may send control signals to any of the 4 front end cards
attached to a given PDC. Each PDC card contains a PCF8574 I$^{2}$C bus expander
chip (by Philips). In conjunction with a decoder, it multiplexes the I$^{2}$C
bus through one of 4 analog switches to one of the front end cards.

\section{System Integration and Operational Characteristics}

\subsection{DC Power Supply System}

The DC power is provided by high current (100 A), well regulated, linear
supplies (Kepco model ATE 6-100M). The supplies are located about 6 meters from
a given ASD card.  The supplies are mounted on the muon support platforms
inside the concrete shielding of the experimental area.  They are remotely
controlled from outside the shield wall by a voltage level.  The supplies are
located as closely as possible to the ASD cards is to minimize ohmic losses and
noise pickup.

The power supplies feed a set of bus bars mounted directly on the support frame
for the chambers. To allow ease in separating the two halves of a given frame,
each half has its own set of copper bus bars, 60 mm$^{2}$ in cross section by
roughly 4~m in length, for each voltage supplied to the ASD cards: $\pm 2.5$ V,
+5 V, GND. In the case of the pad chambers, the -3.5~V power is supplied on a
bus bar of 200~mm$^{2}$ cross section.

\subsection{Oscillation and Noise Control}

A fast high gain amplifier capable of detecting signals in the fC range is
readily susceptible to noise and to feedback. Combating these problems in the
case of the tube ASD card requires three primary measures. First, a copper foil
shields the ends of the anode wires and the ASD inputs. The foil is attached
directly to the ground of the ASD card at one end and to the chamber via
conductive epoxy at the other end. This structure is paramount to providing a
low impedance connection between the ground of the chamber body and the ground
on the ASD card.

The second preventive measure is an rf shield surrounding the card and the
ends of the anode wires. Figure \ref{fig:tubecardmounted} shows the rf shield
with the top cover removed. The shield is made of aluminum. Together with the
chamber body, the rf shield provides essentially $4\pi$ coverage of the tube
readout card and the anode wires.

The third preventive measure is care in routing the signal cabling. These
cables are flat and unshielded, and they have 17 twisted pairs that lead from
the ASD card to the FED modules (see Section \ref{subsec:anodewirefee}). To
minimize the coupling between the signal cable and the ASD inputs, the signal
cables are prevented from touching the chambers or the support frame. To
prevent coupling between the cables, insulating spacers of thickness 5~mm are
inserted between neighboring pairs of cables along the 3~m nearest the ASD
card.

\subsection{Operation and Performance}

Presently, a substantial amount of the muon system is installed. Of the 4
superlayers in the muon system, superlayer 1 is completely in place;
superlayers 3 and 4 each have half a layer in place. The total number of
installed channels is approximately 10800.

Using the existing Front End Drivers and DAQ system, we have measured the noise
performance and detector response at high interaction rates. Typical noise
occupancies for a channel are in the range of a few $\times 10^{-6}$ for
threshold values on the efficiency plateau for our chambers. This measurement
is based on 130000 triggers per channel.

\begin{figure}[h]
   \begin{center}
%   \resizebox{\textwidth}{!}{\includegraphics
%    {occ-rate.eps}} 
\centerline{\epsfxsize 3.5in \epsffile{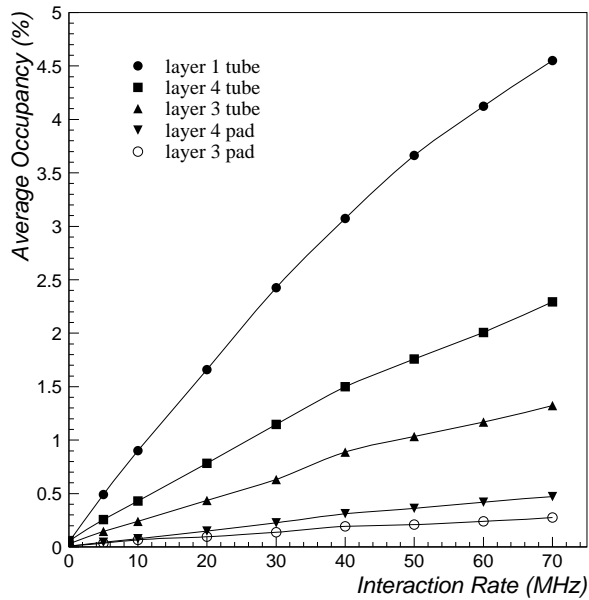}}
%\vspace*{3.5in}
{\caption{\label{fig:occ-rate} \protect \small Average
Occupancy versus Interaction Rate in the Target. The nominal interaction rate
for HERA-B is 40~MHz. The chambers and readout operate well at nearly twice the
nominal interaction rate.}} 
   \end{center}
\end{figure}

We have also measured the occupancy in the chambers when there are interactions
in the HERA-B target. In Figure \ref{fig:occ-rate} is a plot of the average
occupancy in a given superlayer (1, 3, or 4) for a given type of readout
(tube/pad) as a function of the interaction rate (IR) in the target. The
average occupancy scales with the IR and goes toward zero when the IR goes
toward zero. As expected, the rates are highest in the first superlayer since
it is downstream of only part of the muon absorber. The rates are higher in the
anode wires than in the cathode pads because the anode wires detect particles
over a larger area. Occupancies are measured with interaction rates up to 70
MHz, about 1.8 times the rate at which HERA-B will run. Thus, initial
indications are that the 10800 channels currently installed are capable of
operating well at the 40 MHz interaction rate required by HERA-B. The remaining
2/3 of the muon channels will be installed during late 1998 and during 1999.

\section{Acknowledgments}

We wish to thank Albert McGilvra of McGilvra Engineering (Jackson, Michigan)
for the initial design of the microcontroller and power distribution cards. We
also thank Mitch Newcomer of the University of Pennsylvania for arranging for
mass production of the ASD-08 chip and for advice on how to operate it.

\end{document}